# Time evolution of Rényi entropy under the Lindblad equation

Sumiyoshi Abe

*Department of Physical Engineering, Mie University, Mie 514-8507, Japan*
*and Institute of Physics, Kazan Federal University, Kazan 420008, Russia*

**Abstract**  In recent years, the Rényi entropy has repeatedly been discussed for characterization of quantum critical states and entanglement. Here, time evolution of the Rényi entropy is studied. A compact general formula is presented for the lower bound on the entropy rate.

Consider a density matrix $\rho$ that is positive semidefinite and satisfies the normalization condition, $\mathrm{tr}\,\rho = 1$. Its Rényi entropy defined by

$$S_\alpha = \frac{1}{1-\alpha} \ln \mathrm{tr}\,\rho^\alpha \qquad (\alpha > 0) \qquad (1)$$

provides a useful tool for characterizing entanglement contained in $\rho$ (see Refs. [1-3], for example). This generalized entropy obviously converges to the von Neumann entropy, $S = -\mathrm{tr}(\rho \ln \rho)$, in the limit $\alpha \to 1$. $\rho$ may instantaneously be diagonalized in a certain orthonormal basis $\{|u_i\rangle\}_i$ as

$$\rho = \sum_i p_i |u_i\rangle\langle u_i|, \qquad (2)$$

where the eigenvalues, $p_i$'s, are in the range $[0,1]$ and satisfy $\sum_i p_i = 1$. In this form of $\rho$, the Rényi entropy is written as $S_\alpha = (1-\alpha)^{-1} \ln\left(\sum_i p_i^\alpha\right)$, which is analogous to its classical definition [4].

Some comments are in order. Firstly, the Rényi entropy is not uniformly continuous as a functional unless $\alpha \to 1$. This has been discussed in Ref. [5], where the concept of so-called "Lesche stability" has been introduced. Later, relevant issues have further been studied in Ref. [6] for other generalized entropies. Problems regarding uniform continuity of a functional usually matter in the neighborhoods of the boundary of its domain. Therefore, one would ignore the problems to use the Rényi entropy or other similar generalized entropies to extend traditional thermostatistics. Then, the second comment comprises the fact that, like other generalized entropies, discussions of this kind, however, are illegitimate due to the presence of unwarranted biases in such generalized maximum entropy schemes [7]. In addition, the Rényi entropy does not satisfy the concept of subset independence that is the fourth of the Shore-Johnson axioms [8] for the maximum entropy principle.

In this article, we discuss the dynamical behavior of the Rényi entropy by employing a master equation of the Lindblad type [9,10]. We present a compact formula for the lower bound on the entropy rate, which generalizes the result known for the von Neumann entropy [11] (see also Ref. [12]).

Let us recall the Lindblad equation for a density matrix $\rho$ [9,10]

$$i\frac{\partial \rho}{\partial t} = [H, \rho] - \frac{i}{2} \sum_n c_n \left( L_n^\dagger L_n \rho + \rho L_n^\dagger L_n - 2 L_n \rho L_n^\dagger \right),  \qquad (3)$$

where $\hbar$ is set equal to unity. $H$ stands for a system Hamiltonian and $L_n$'s are referred to as the Lindbladian operators that describe interaction between the system and its environment. $c_n$'s are $c$ numbers and have to be nonnegative in order for the density matrix to remain positive semidefinite in the course of time evolution. This is the most general quantum master equation that is linear, Markovian, and positive semidefiniteness preserving.

Now, the time derivative of the Rényi entropy in Eq. (1) is written as follows:

$$\frac{dS_\alpha}{dt} = \sum_n c_n \Gamma_n, \qquad (4)$$

where $\Gamma_n$ is given by

$$\Gamma_n = \frac{\alpha}{1-\alpha} \frac{1}{\mathrm{tr}\rho^\alpha} \mathrm{tr}\left(\rho^{\alpha-1} L_n \rho L_n^\dagger - \rho^\alpha L_n^\dagger L_n\right), \qquad (5)$$

provided that the identical relation, $\mathrm{tr}(\rho^{\alpha-1}[H,\rho]) = 0$, has been used. The presence of $\rho^{\alpha-1}$ in Eq. (5) mathematically forces us to assume in the case $\alpha < 1$ positive definiteness of the density matrix rather than positive semidefiniteness, but the final results will be free from this problem.

In what follows, we show that the quantity in Eq. (5) satisfies

$$\Gamma_n > \left\langle [L_n^\dagger, L_n] \right\rangle_\alpha, \qquad (6)$$

where $\langle A \rangle_\alpha$ is defined by

$$\langle A \rangle_\alpha = \frac{\mathrm{tr}(A\rho^\alpha)}{\mathrm{tr}\rho^\alpha}, \qquad (7)$$

which is referred to as the $\alpha$ average of $A$. The instantaneous diagonalization in Eq. (2) makes the quantity in Eq. (7) be rewritten as $\langle A \rangle_\alpha = \sum_i P_i^{(\alpha)} \langle u_i | A | u_i \rangle$, where $P_i^{(\alpha)} \equiv p_i^\alpha / \sum_j p_j^\alpha$ is formally equivalent to the escort distribution associated with $p_i$ [13].

Before proceeding, here we wish to point out the fact that, like the Rényi entropy itself, the $\alpha$-average is not Lesche-stable either [14] under deformation of $\rho$ unless $\alpha \to 1$. In addition, such a generalized average is not consistent with the principles of quantum mechanics [15,16]. However, this issue is irrelevant to our discussion, since we are regarding the $\alpha$ average as a purely mathematical expression, here.

Now, let us show that Eq. (6) holds. The proof is elementary.

Substituting Eq. (2) into Eq. (5), we have

$$\Gamma_n = \frac{\alpha}{1-\alpha} \frac{1}{\operatorname{tr}\rho^\alpha} \left\{ \sum_{i,j} p_i^{\alpha-1} p_j \left|\langle u_i | L_n | u_j \rangle\right|^2 - \sum_i p_i^\alpha \langle u_i | L_n^\dagger L_n | u_i \rangle \right\}. \qquad (8)$$

We wish to replace the double summation inside the braces on the right-hand side with a single summation. For this purpose, it is necessary to separately examine two cases: (i) $0 < \alpha < 1$ and (ii) $\alpha > 1$. Let us consider $f(x) = x^\alpha$ of positive $x$. This function is concave in case (i) and convex in case (ii). Therefore, we have [17]

$$\text{(i)} \; \left[\lambda p_i + (1-\lambda) p_j\right]^\alpha > \lambda p_i^\alpha + (1-\lambda) p_j^\alpha,$$

$$\text{(ii)} \; \left[\lambda p_i + (1-\lambda) p_j\right]^\alpha < \lambda p_i^\alpha + (1-\lambda) p_j^\alpha, \qquad (9)$$

where $\lambda \in (0,1)$. On the other hand, since it is true for any positive $y$ [17] that, (i) $y^\alpha - 1 \leq \alpha(y-1)$, (ii) $y^\alpha - 1 \geq \alpha(y-1)$; with the equalities being for $y=1$, it follows that

$$\text{(i)} \; p_i^\alpha \left[\lambda + (1-\lambda)\frac{p_j}{p_i}\right]^\alpha \leq p_i^\alpha \left\{ \alpha \left[\lambda + (1-\lambda)\frac{p_j}{p_i} - 1\right] + 1 \right\},$$

(ii) $p_i^\alpha \left[ \lambda + (1-\lambda)\frac{p_j}{p_i} \right]^\alpha \geq p_i^\alpha \left\{ \alpha \left[ \lambda + (1-\lambda)\frac{p_j}{p_i} - 1 \right] + 1 \right\}.$ (10)

Combining Eq. (9) with Eq. (10), we have

(i) $p_i^\alpha \left\{ \alpha \left[ \lambda + (1-\lambda)\frac{p_j}{p_i} - 1 \right] + 1 \right\} > \lambda p_i^\alpha + (1-\lambda) p_j^\alpha,$

(ii) $p_i^\alpha \left\{ \alpha \left[ \lambda + (1-\lambda)\frac{p_j}{p_i} - 1 \right] + 1 \right\} < \lambda p_i^\alpha + (1-\lambda) p_j^\alpha.$ (11)

It may be of interest to observe that $\lambda$ disappears from Eq. (11). Thus we obtain

(i) $\alpha p_i^{\alpha-1} p_j > (\alpha-1) p_i^\alpha + p_j^\alpha,$ (ii) $\alpha p_i^{\alpha-1} p_j < (\alpha-1) p_i^\alpha + p_j^\alpha.$ (12)

Therefore, we find that, in both cases (i) and (ii), Eq. (8) satisfies the following inequality:

$$\Gamma_n > \frac{1}{\mathrm{tr}\,\rho^\alpha} \sum_i p_i^\alpha \langle u_i | [L_n^\dagger, L_n] | u_i \rangle,$$ (13)

which proves Eq. (6).

An immediate consequence from the results in Eqs. (4) and (6) is that if $L_n$'s are normal [18], that is, $[L_n^\dagger, L_n] = 0$, then the Rényi entropy rate is positive. In addition, the result clearly reproduces the one given in Ref. [11] in the limit $\alpha \to 1$.

In conclusion, we have derived a compact formula for the lower bound on the time derivative of the Rényi entropy under the Lindblad equation. The result is expected to be useful for studying dynamics of quantum entanglement in the Markovian approximation. However, there is no *a priori* reason for dynamics of a subsystem, i.e., subdynamics, to be Markovian [19]: if the total system is in a strongly entangled state, then subdynamics generically fails to be Markovian and even the subsystem Hamiltonian may not exist. It is therefore of obvious importance to further generalize the present discussion to non-Markovian dynamics.

This work was supported in part by a Grant-in-Aid for Scientific Research from the Japan Society for the Promotion of Science (No. 26400391) and by the Program of Competitive Growth of Kazan Federal University from the Ministry of Education and Science of the Russian Federation.

______________________

[1]   B.-Q. Jin and V. E. Korepin, J. Stat. Phys. **116**, 79 (2004).

[2]   F. Ares, J. G. Esteve, and F. Falceto, Int. J. Geom. Methods Mod. Phys. **12**, 1560002 (2015).

[3]   A. Hamma, S. M. Giampaolo, and F. Illuminati,

Phys. Rev. A **93**, 012303 (2016).

[4]   A. Rényi, *Probability Theory* (North-Holland, Amsterdam, 1970).

[5]   B. Lesche, J. Stat. Phys. **27**, 419 (1982).

[6]   S. Abe, Phys. Rev. E **66**, 046134 (2002).

[7]   S. Pressé, K. Ghosh, J. Lee, and K. A. Dill, Phys. Rev. Lett. **111**, 180604 (2013); S. Pressé, K. Ghosh, J. Lee, and K. A. Dill, Entropy **17**, 5043 (2015).

[8]   J. E. Shore and R. W. Johnson, IEEE trans. Inform. Theory **26**, 26 (1980); **29**, 942 (1983).

[9]   G. Lindblad, Commun. Math. Phys. **48**, 119 (1976).

[10]  V. Gorini, A. Kossakowski, and E. C. G. Sudarshan, J. Math. Phys. **17**, 821 (1976).

[11]  F. Benatti and H. Narnhofer, Lett. Math. Phys. **15**, 325 (1988).

[12]  C. Ou, R. V. Chamberlin, and S. Abe, e-print 1601.07874.

[13]  R. Bhatia, *Matrix Analysis* (Springer-Verlag, New York, 1997).

[14]  C. Beck and F. Schlögl, *Thermodynamics of Chaotic Systems: An Introduction* (Cambridge University Press, Cambridge, 1993).

[15]  S. Abe, EPL **84**, 60006 (2008); S. Abe, J. Stat. Mech. P07027 (2009).

[16]  L. D. Landau and I. M. Lifshitz, *Quantum Mechanics* (Pergamon, Oxford, 1958).

[17]  S. Abe, J. Phys.: Conf. Ser. **394**, 012003 (2012).

[18]  G. H. Hardy, J. E. Littlewood and G. Pólya, *Inequalities*, 2nd ed. (Cambridge University Press, Cambridge, 1952).

[19]  H.-P. Breuer, E.-M. Laine, J. Piilo, and B. Vacchini, Rev. Mod. Phys. **88**, 021002 (2016).